\documentclass[11pt,a4paper]{article}

\usepackage{cite}

\usepackage{multicol,bbm}

\usepackage{amsmath, amssymb, amsfonts}

\usepackage{graphicx}
\usepackage{color}

\makeatletter \@addtoreset{equation}{section} \makeatother

\begin{document}

\title{Coupled fermion-kink system in Jackiw-Rebbi model}

\author{A. Amado$^1$\thanks{andreamado@df.ufpe.br}\ \ and A. Mohammadi$^2$\thanks{a.mohammadi@fisica.ufpb.br}\\
\textit{$^1$Departamento de F\'isica, Universidade Federal de Pernambuco,}\\\textit{52171-900 Recife, PE, Brazil}\vspace{0.3cm}\\
\textit{$^2$Departamento de F\'{\i}sica, Universidade Federal de Campina Grande}\\\textit{
Caixa Postal 10071, 58429-900 Campina Grande, Para\'{\i}ba, Brazil}}

\maketitle

\begin{abstract}
In this paper we study Jackiw-Rebbi model, in which a massless fermion is coupled to the kink of $\lambda \phi^4$ theory through a Yukawa interaction. In the original Jackiw-Rebbi model the soliton is prescribed. However, we are interested in the back-reaction of the fermion on the soliton besides the effect of the soliton on the fermion. 
Also, as a particular example, we consider a minimal supersymmetric kink model, $\mathcal{N}=1$, in ($1+1$) dimensions. In this case, the bosonic self-coupling, $\lambda$, and the Yukawa coupling between fermion and soliton, $g$, have specific relation, $g=\sqrt{\lambda/2}$.
As the set of coupled equations of motion of the system is not analytically solvable, we use a numerical method to solve it self-consistently. We obtain the bound energy spectrum, bound states of the system and the corresponding shape of the soliton using a relaxation method, except for the zero mode fermionic state and threshold energies which are analytically solvable. With the aid of these results we are able to show how the soliton is affected in general and supersymmetric cases. The results we obtain are consistent with the ones in the literature, considering the soliton as background.
\end{abstract}

\maketitle

\section{Introduction}

Solitons named by Zabusky and Kruskal in 1965 \cite{ZK}, first appeared as a solution for KdV equation \cite{KdV}. They play important roles in diverse areas of physics, biology and engineering \cite{Rajaraman,Drazin,Manton}. In one spatial dimension kinks, and in higher spatial dimensions vortices,  monopoles, instantons and domain walls,  are amongst the most important ones in this category. They are topological configurations which appear in different areas of physics such as high energy, atomic and condensed matter physics \cite{CM1,CM2,CM3,CM4,CM5}. These topologically nontrivial configurations cannot be continuously deformed into a trivial vacuum configuration.

The coupling of the fermionic field to other fields alters the energy spectrum of the fermionic field as well as the wave function which can cause many interesting phenomena. 
Models including coupled fermionic and bosonic fields are crucial in many branches of physics, specially when the bosonic field has the form of a soliton. As solitons can be viewed as extended particles with finite mass, i.e. finite energy at rest, the systems consisting of coupled fermionic and solitonic fields can be good candidates to describe extended objects such as hadrons. Since 1958, with Skyrme's pioneering works \cite{Skyrme1,Skyrme2,Skyrme3,Skyrme4,Skyrme5}, many physicists have tried to explain the hadrons and their strong interactions nonperturbatively using phenomenological nonlinear field theories \cite{qcd1,qcd2,qcd3}. 

The presence of a soliton in fermion-soliton systems distorts the fermion vacuum state and consequently can induce nonzero vacuum polarization and Casimir energy. The corresponding nontrivial topology induces nonzero vacuum expectation values for physical observables. 
These phenomena have been widely discussed in the literature for different types of solitons and in different dimensions (e.g. \cite{Azi4,Azi3,Farid2,Mello1,Azi1,Azi2}). Moreover,
when a fermion interacts with a soliton, an interesting phenomenon 
 occurs which is the assignment of fractional fermion number to the solitonic state. 
In a theory where all the fields carry integer quantum numbers, the emergence of fractional quantum numbers has attracted a lot of interest. 
Jackiw and Rebbi pointed out the occurrence of the fractional fermion
number for the first time \cite{jackiw1976}.
Much of the work in this area has been inspired by the Jackiw and Rebbi's pioneering work. They considered some models with the charge conjugation symmetry where the fermion is coupled to a bosonic background field in the form of a soliton.
They have shown that the existence of a fermionic nondegenerate zero mode implies the soliton with fermion number one half.

Coupled fermion-soliton systems also appear in the braneworld scenarios in the context of the localization of the Standard Model fields
on the brane. Localizing a fermion on the brane was first described by the original work \cite{Shaposhnikov}, where our Universe can be realized inside a domain wall embedded in a (4+1)-dimensional world.   
The localization of spin 1/2 fermions on thin branes due to a soliton, is performed via the mechanism introduced by Jackiw and Rebbi \cite{jackiw1976} originally to demonstrate the fermion charge-fractionization phenomenon. 
Also, higher-dimension extensions for this mechanism have been studied in literature \cite{extension1,extension2,extension3,extension4}. 
Besides that, the localization of fermions on a double-brane in warped space-time has been studied (e.g. \cite{double_brane1}).

In most of the models  investigated in the literature consisting of coupled fermion-soliton systems, the soliton is considered as a background field. The main reason is that solving the nonlinear system treating both fields as dynamical\footnote{In this paper we use the word dynamical, in contrast to prescribed, to refer to the result of the equations of motion considering both fermionic and bosonic fields together. In the prescribed model we consider the soliton to be a prescribed (or background) field that does not receive any backreaction from the fermion.} is in general extremely difficult analytically \cite{Leila}. In principle, the soliton in these systems can have infinitely different shapes. Based on Jackiw and Rebbi's work \cite{jackiw1976}, the back-reaction of the fermion on the soliton is small when the coupling of the interaction term is small. Therefore, in this regime considering the soliton to be a prescribed field is a good approximation, although when the coupling is not small it may fail considerably. 
In Jackiw-Rebbi model, the system has charge and particle conjugation symmetries. The particle conjugation symmetry relates each fermionic mode with energy $E$ to the one with energy $-E$, which makes the energy spectrum to be completely symmetric with respect to the line $E=0$. 

In this paper we study the Jackiw-Rebbi model with a massless fermion coupled to the kink of $\lambda \phi^4$ theory as well as a minimal supersymmetric kink model, as an example, and solve the system using a numerical method. As we solve the system self-consistently, it is possible to analyze not only the effect of the soliton on the fermion field but also the backreaction of the fermion field on the soliton. 
In this case, the non-degenerate zero mode is always present regardless of the values of the parameters of the model (the same happens in Jackiw-Rebbi model), and surprisingly the soliton receives no back-reaction due to this mode.
In the dynamical model, the system loses the charge and particle conjugation symmetries, resulting in a nonsymmetrical energy spectrum with respect to the $E=0$ line. We show that our results are consistent with the ones discussed in literature with the prescribed soliton, where the system recovers its charge and particle conjugation symmetries.

This paper is organized in five sections: in section 2 we briefly introduce the fermion-kink model in ($1+1$) dimensions as well as the formulation of the problem. In this section we write the lagrangian describing our model, in components, and the resultant equations of motion. In section 3 we obtain the fermionic bound states and bound energies of the system. We find the fermionic zero mode and threshold bound energies analytically. To obtain the other fermionic bound energies, bound states and the shape of the soliton we use a numerical method called relaxation method. At the end of the section we show the classical soliton mass as well as the back-reaction of the fermion on the soliton. In section 4 we solve the system for the particular case of the supersymmetric kink with $\mathcal{N}=1$, as an example. Finally, section 5 is devoted to summarize and discuss our results.

\section{Fermion-kink system}
We consider a fermion-soliton system in (1+1) dimensions given by the following lagrangian
\begin{align}\label{lagrangian_final1}
	\mathcal{L} = \frac{1}{2} \partial_\mu\phi\partial^\mu \phi
	+ \frac{1}{2}\bar{\psi}\,i\gamma^\mu\partial_\mu\psi
	- g\,\phi\,\bar{\psi}\psi
	- V(\phi),
\end{align}
with $V(\phi)=\frac{\lambda}{4} \left( \phi^2 - \frac{m^2}{\lambda} \right)^2$, the known $\phi^4$ theory potential. In order to guarantee a well-defined energy for the soliton, we have $\phi(x\to\pm\infty)=\pm\frac{m}{\sqrt{\lambda}}\equiv\pm\phi_0$.
Considering the field $\phi$ to be static, the Euler-Lagrange equations of the system are given by 
\begin{align}\label{equations_of_motion_1}
     -i\gamma_\mu\partial^\mu \psi + 2 g \,\phi\,\psi &= 0,\nonumber\\
    \phi'' - \lambda\phi\left(\phi^2-\frac{m^2}{\lambda}\right) - g \bar{\psi}\psi &= 0,
\end{align}
where prime denotes differentiation with respect to $x$. Defining $\chi \equiv \phi/\phi_0$ and $\psi =e^{-i E t} \begin{pmatrix}
		\psi_1 + i\, \psi_3 \\
		\psi_2 + i\, \psi_4
	   \end{pmatrix} $, these equations become
\begin{align}\label{equations_of_motion_diff_coupling}
     E\ \psi_1 + \psi_2' - 2 g \, \phi_0\,\chi\,\psi_2 &= 0,\nonumber\\
     E\ \psi_2 - \psi_1' - 2 g \, \phi_0\,\chi\,\psi_1 &= 0,\nonumber\\
     E\ \psi_3 + \psi_4' - 2 g \, \phi_0\,\chi\,\psi_4 &= 0,\nonumber\\
     E\ \psi_4 - \psi_3' - 2 g \, \phi_0\,\chi\,\psi_3 &= 0,\nonumber\\
    -\chi'' + m^2\chi\left(\chi^2-1\right) + 2 g/\phi_0 \left(\psi_1\psi_2 + \psi_3\psi_4\right) &= 0.
\end{align}
in which the representation for the Dirac matrices is chosen as $\gamma^0 = \sigma_1$, $\gamma^1 = i \sigma_3 $ and $\gamma^5 = \sigma_2$. 
Due to the symmetry in the equation system, there are only two independent degrees of freedom in $\psi$. Therefore, we can solve only the real part of the spinor field, components $\psi_1$ and $\psi_2$. Also, we rescale all the quantities to dimensionless ones as $\psi\to\sqrt{m}\psi$, $\chi\to\chi$, $E\to mE$, $\phi_0 \to \phi_0$ ($\lambda\to m^2\lambda$), $g\to m g$ and $x\to x/m$ \footnote{We write the mass dimension in parenthesis when deemed necessary in order to avoid confusion.}.
	   
As can be seen in the lagragian (\ref{lagrangian_final1}), the fermion field interacts nonlinearly with the pseudoscalar field. The system cannot be solved analytically without imposing the soliton to be a background field. Thus, using a numerical method we solve this coupled set of differential equations self-consistently and find the fermionic bound states and bound energies as well as the shape of the soliton.
The shape of the static soliton in the model we consider here is not prescribed and is determined by the equations of motion.
With this, besides the effect of the soliton on the fermion, we are able to obtain at the classical level the effect of the fermion on the soliton (the back-reaction), within our numerical restrictions.
The main advantage is to help us understand the system beyond the regime considered in the literature where the soliton can be treated as background which is equivalent to $g/\phi_0\to0$ limit.

In the limit $g/\phi_0\to0$ ($g\to0$ and/or $\lambda\to0$), the last equation in (\ref{equations_of_motion_diff_coupling}) decouples from the others and has analytical solution, i.e. kink of $\lambda\, \phi^4$ theory. The solutions of some analogous systems in this limit have been studied in detail in \cite{jackiw1976,farid}. In this limit, the solutions for this equation are
\begin{align}\label{chi_static}
	\chi_{bg}(x) =\pm \tanh \left( \frac{x-x_0}{\sqrt{2}} \right).
\end{align}
In this paper we consider the positive sign in eq. (\ref{chi_static}).
One can calculate the classical soliton mass using the expression
\begin{align}\label{eq:soliton_mass}
	M_{cl} = \int_{-\infty}^{\infty} \left(\frac{1}{2} (\phi_0\chi)'^2 + V(\phi_0\chi)\right) \mbox{d}x,
\end{align}
that is $M_{cl} = \frac{2 \sqrt{2}}{3}\,m\,\phi_0^2$ for the kink of $\lambda\, \phi^4$ theory.
We solve the set of coupled equations (\ref{equations_of_motion_diff_coupling}) self-consistently and check the results with the ones considering the soliton as background.

In $g/\phi_0\to0$ limit, the system has charge and particle conjugation symmetries. In this case, the charge conjugation operator is $\gamma^1$, which is also the particle conjugation operator. This means that the negative and positive energy spectrums are mirror images of each other around $E=0$, in this limit. However, as the ratio $g/\phi_0$ increases, the $g\bar{\psi}\psi$ term in second equation of (\ref{equations_of_motion_1}) cannot be neglected anymore and breaks these symmetries. As a result, one can see that the energy spectrum is not symmetric around $E=0$ in general. 
In \cite{farid} the authors have obtained a symmetric spectrum around $E=0$. However, in the system they have considered, the soliton is a background field no matter how big the ratio $g/\phi_0$ could be. Therefore, the result obtained in the referred paper is not a good approximation for the model described by the lagrangian (\ref{lagrangian_final1}) when the ratio $g/\phi_0$ is not small enough to consider the prescribed soliton. In this paper we compare our results with the ones with background soliton. Although the term $g\bar{\psi}\psi$ in the equation of motion breaks the charge and particle conjugation symmetries, it does not break parity symmetry. This way, the system has parity and as a result the wavefunctions display this symmetry regardless of the value of $g/\phi_0$. This feature can be seen shortly in our numerical results.

\section{Bound states and bound energies}
We obtain the zero energy bound state and threshold energies analytically, although to find the other bound states we have to rely on a numerical method.

\subsection{Zero energy bound state}
For this state the equations of motion are simplified and we are able to obtain the analytical solution of the system in the whole $g$ and $\phi_0$ intervals.
Taking $E = 0$, the equation system becomes
\begin{align}\label{a14}
      \psi_1' + 2 g \phi_0\ \chi\ \psi_1 &= 0,\nonumber\\
      \psi_2' - 2 g \phi_0\ \chi\ \psi_2 &= 0,\nonumber\\
    -\chi'' + \chi\left(\chi^2- 1\right) + 2 g /\phi_0 \ \psi_1\psi_2 &= 0.
\end{align}
It turns out that the first two equations can be easily solved as functions of $\chi$, yielding
\begin{align}\label{a15}
     \psi_1(x) &= a_1\,\mbox{e}^{-2 g \phi_0\int_1^x\, \chi (x')\, dx'},\nonumber\\
     \psi_2(x) &= a_2\,\mbox{e}^{2 g \phi_0\int_1^x\, \chi (x')\, dx'}.
\end{align}
Notice that if we define $f(x) \equiv \mbox{e}^{-2 g \phi_0\int_1^x \, \chi (x')\, dx'}$, either $f(x\to\pm\infty) = 0$ and consequently $f^{-1}(x\to\pm\infty)$ diverges or $f^{-1}(x\to\pm\infty) = 0$ and as a result $f(x\to\pm\infty)$ diverges. As the normalization of the divergent components should be zero, we can conclude either $a_1 = 0$ or $a_2 = 0$. Thus, the term with $\psi$-dependence in the last equation vanishes and we find
\begin{align}\label{a16}
	    -\chi'' + \chi\left(\chi^2- 1\right) = 0,
\end{align}
which corresponds to the trivial kink equation with the known solution eq. (\ref{chi_static}).
This way one can determine $f(x) = \mbox{e}^{-2 g \phi_0\,\int_1^x  \tanh\left[(x'-x_0)/\sqrt{2}\right] \, d x'} \propto \left[\cosh\left(\frac{x-x_0}{\sqrt{2}}\right)\right]^{-2\sqrt{2}g \phi_0}$. Requiring the wave function to be normalized, we obtain 
\begin{align}\label{a18}
	\psi(x) = \sqrt{\frac{\Gamma(1/2+2\sqrt{2} g \phi_0)}{\sqrt{2\pi}\,\Gamma(2\sqrt{2} g \phi_0)}}
			\begin{pmatrix}
				\left[\cosh\left(\frac{x-x_0}{\sqrt{2}}\right)\right]^{-2\sqrt{2}g \phi_0} \\
				0
		  	\end{pmatrix}.
\end{align}
It is important to notice that based on this result, the back-reaction of the fermion on the soliton is zero for the fermionic zero mode. 
If a system does not possess this symmetry, the back-reaction for the fermionic zero mode is nonzero in general \cite{Leila}. 
Although the system considered here does not respect this symmetry, the zero mode still does not receive back-reaction. This happens because the term $g\bar \psi\psi$ is null for the zero mode and the system recovers particle conjugation symmetry.

\subsection{Threshold states}
\label{sec:threshold}
Threshold or half-bound states are the states where the fermion field goes to a constant at spatial infinity. For these states when $x\to\infty$ the wave function is finite, but does not decay fast enough to be square-integrable \cite{Graham,Dong1,Dong2}.

To find such states in our system we solve the system of equations at $x\to\pm\infty$. We write $\psi$ in the form $\psi(x\to\pm\infty) = e^{-i E t} \begin{pmatrix}
							c_1 \\
							c_2
				   	  \end{pmatrix}$, where $c_i$'s are arbitrary constants.

Applying the conditions $\chi(x\to\pm\infty) = \pm1$, $\chi'(x\to\pm\infty) = \chi''(x\to\pm\infty) = 0$ and $\psi'(x\to\pm\infty) = \begin{pmatrix}0 \\0\end{pmatrix}$ to the set of equations of motion, we obtain
\begin{align}\label{threshold12}
     E\,c_1  - 2 g \phi_0\,\chi\,c_2 &= 0,\nonumber\\
     E\,c_2 -  2 g \phi_0\,\chi\,c_1 &= 0,\nonumber\\
     \frac{2g}{\phi_0}c_1c_2 &= 0.
\end{align}
This set of equations has nontrivial solution only when the last equation decouples, i.e. $g/\phi_0\to0$. Solving the first two equations in this case, it is easy to show that the energies of the threshold states are $E = \pm2 g \phi_0$, as expected.

\subsection{Numerical method}

The remaining bound states cannot be found analytically and a numerical method is required. We use a relaxation method that starts with an initial guess and iteratively converges to the solution of the system.
We start with the known energy spectrum and bound states where the soliton is a background field \cite{farid} and find the solution of the system, considering $\phi_0=\pi$ which gives a soliton with winding number one. There are two first order and one second order differential equations (eq. (\ref{equations_of_motion_diff_coupling})). The latter can be transformed to a set of two first order equations as 
\begin{align}\label{a20}
    p \equiv \chi', \qquad -p' + \chi\left(\chi^2- 1\right) + \frac{2g}{\phi_0}\psi_1\psi_2= 0. 
\end{align}
To find the fermion energy eigenvalue we also introduce the equation $E' = 0$, reflecting the fact that energy is constant. 
Moreover, we fix the translational symmetry of the system by choosing $x_0 = 0$. 

\begin{figure}[h]
  \centering
  \fbox{\begin{tabular}{c c c}
\includegraphics[width=0.31\textwidth]{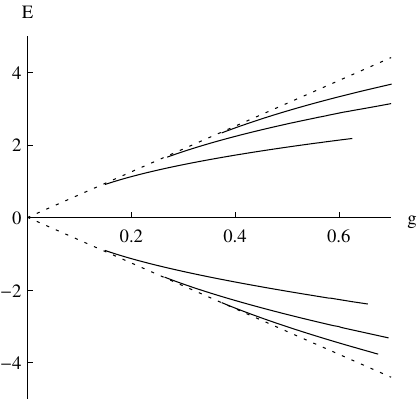} & \includegraphics[width=0.29\textwidth]{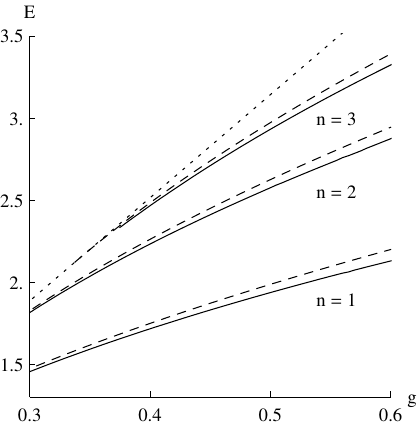} & \includegraphics[width=0.29\textwidth]{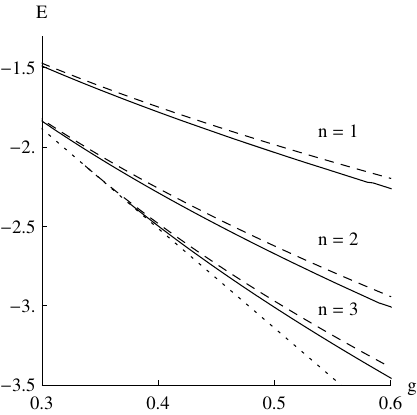} 
\end{tabular}}
\caption{The fermionic bound energies as a function of $g$, for the soliton with winding number one. Left graph: solid curves depict the positive and negative fermionic bound energies, in the dynamical model.  Middle graph: solid and dashed curves depict positive fermionic bound energies in the dynamical and background model, respectively. Right graph: solid and dashed curves depict negative fermionic bound energies in the dynamical and background model, respectively. In all graphs dotted lines depict the fermionic threshold energies in $g/\phi_0\to0$ limit.}
  \label{fig:spectrum}
\end{figure}

Now, there are five coupled first-order differential equations which need five boundary conditions. Among the several boundary conditions available, we choose the following
\begin{align}\label{a21}
    \chi(\pm \infty) = \pm 1, \quad
    \chi(0) = 0, \quad
    \psi_1(\pm\infty) = 0.
\end{align}
In Fig. (\ref{fig:spectrum}) we show the energy spectrum as a function of the coupling $g$ for the soliton with winding number one, $\phi_0=\pi$. The left graph shows the first three positive and negative energy levels of the system. In the middle and right graphs, the positive and negative energy levels, respectively, are zoomed in specific regions of the parameters. We have depicted our result with the solid curves and compared with the dashed ones, the background soliton results.
 As one can see, our results and the ones with the background kink become more different as $g$ grows. Also, the symmetry of the energy levels around the line $E=0$ which was expected in the background model, breaks gradually when we increase $g$ from zero. 
 This becomes evident noticing that both positive and negative energies are lower than their counterparts in the background model.

\begin{figure}[h]
\begin{center}
	\fbox{\includegraphics[width=0.8\textwidth]{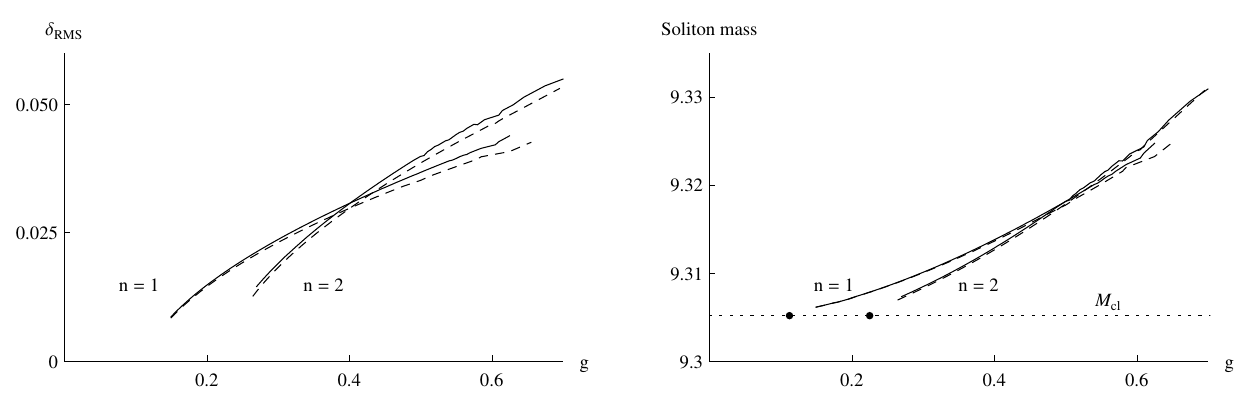}}
	\caption{Left graph: $\delta_{RMS}$ as a function of the coupling $g$. The solid curves denote the $\delta_{RMS}$ for the first two positive energy levels and dashed ones for the first two negative energy levels. Right graph: Soliton mass as a function of the coupling $g$. The solid curves depict the soliton mass for the first two positive energy levels and the dashed ones for the first two negative energy levels. The dotted line corresponds to the soliton mass in the background model. The two black dots are the points with specific values of $g$ for which the represented bound states first appear.}
    \label{fig:dRMS_mass_vs_g}
\end{center}
\end{figure}

As a measure of the back-reaction of the fermion on the soliton we calculate the root mean squared deviation between the soliton in the dynamical model and the background one, $\delta_{RMS}$.
In the left graph of Fig.\,\ref{fig:dRMS_mass_vs_g}, one can observe that the backreaction increases with the coupling $g$. In small values of this coupling, the difference in the backreaction for positive and negative energy levels is low and as $g$ grows this difference increases. It reflects the fact that in $g/\phi_0\to0$ limit the particle conjugation symmetry is present, being gradually broken with increasing $g$ which distorts the symmetry between the positive and negative energy levels.

As a final result, we show the soliton mass as a function of the coupling $g$ in the right graph of Fig.\,\ref{fig:dRMS_mass_vs_g}. We can see that by increasing the coupling $g$, the soliton mass starts diverging from the classical result, as expected.

\section{An example: supersymmetric kink model}
We consider the minimal supersymmetry, $\mathcal{N}=1$, in a (1+1) dimensional field theory. 
The supersymmetric lagrangian has the form \cite{Shifman:1998zy, Morteza}
\begin{align}\label{a6}
	\mathcal{L} = \frac{1}{2}\left( \partial_\mu \phi\partial^\mu \phi +\bar{\psi}\ i\gamma^\mu\partial_\mu \psi + \mbox{F}^2 \right) + W_\phi\ \mbox{F}\ - \frac{1}{2}W_{\phi\phi}\bar{\psi}\psi,
\end{align}
where the subscript $\phi$ shows the derivative with respect to $\phi$ and $F$ is an auxilary field. Using the Euler-Lagrange equations and choosing the bosonic potential to be the kink potential $V(\phi)= \frac{\lambda}{4} \left(\phi^2 - \frac{m^2}{\lambda}\right)^2$, i.e. $V(\phi)=\frac{1}{2}W_{\phi}^2$, we obtain
\begin{align}\label{lagrangian_final}
	\mathcal{L}_{kink} = \frac{1}{2} \partial_x\phi\partial^x \phi
	+ \frac{1}{2}\bar{\psi}\,i\gamma^\mu\partial_\mu\psi
	- \sqrt{\frac{\lambda}{2}} \,\phi\,\bar{\psi}\psi
	- \frac{\lambda}{4} \left( \phi^2 - \frac{m^2}{\lambda} \right)^2,
\end{align}
in which $\phi$ is considered static.
Note that the supersymmetry relates the bosonic self-coupling $\lambda$ and the Yukawa interaction coupling $g$. 

Taking $g=\sqrt{\lambda/2}$ in the set of equations of motion (\ref{equations_of_motion_diff_coupling}) and considering only two independent degrees of freedom in $\psi$, $\psi_1$ and $\psi_2$, we obtain
\begin{align}\label{equations_of_motion}
     E\ \psi_1 + \psi_2' - \sqrt{2} \,\chi\,\psi_2 &= 0,\nonumber\\
     E\ \psi_2 - \psi_1' - \sqrt{2} \,\chi\,\psi_1 &= 0,\nonumber\\
    -\chi'' + \chi\left(\chi^2-1\right) + \sqrt{2}\lambda \psi_1\psi_2 &= 0,
\end{align}
In $g/\phi_0\to0$ limit which is equivalent to $\lambda\to0$ ($\phi_0\to\infty$) in the supersymmetric case, there are three fermionic bound states with energies $0$ and $\pm\sqrt{3/2}\,(m)$ and two threshold ones with energies $E=\pm\sqrt{2}\,(m)$ \cite{Morteza,farid}. 

We solve the system dynamically and discuss the case where the soliton can be considered as background as well.
Again the zero energy bound state and threshold energies can be obtained analytically, although to find the other bound states we have to solve the system numerically.

For the zero mode the equations of motion are simplified and we are able to obtain the analytical solution of the system in the whole $\phi_0$ interval. 
Using the same method as before we observe that, interestingly, the solutions show to be $\phi_0$ independent. Requiring the wave function to be normalized, we obtain 
\begin{align}\label{a18}
	\psi(x) = \sqrt{\frac{3}{4\sqrt{2}}}
			\begin{pmatrix}
				\text{sech}^{2}\left(\frac{x-x_0}{\sqrt{2}}\right) \\
				0
		  	\end{pmatrix}.
\end{align}
One can use the same method to find the threshold states which gives $E = \pm\sqrt{2}\,(m)$.
 
To find the other bound states and the corresponding parameters of the system we start with the known energy spectrum and bound states in $\lambda\to0$ limit and solve the set of equations for the whole region of $\phi_0$ within the numerical restrictions. The same boundary conditions as in (\ref{a21}) are considered.

\begin{figure}[h]
\begin{center}
	\fbox{\includegraphics[width=0.8\textwidth]{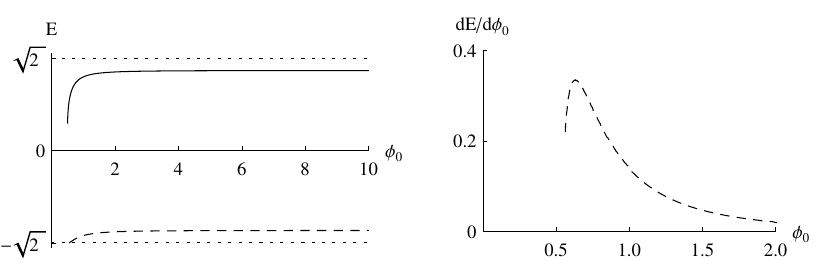}}
	\caption{Left graph: The fermionic bound energies as a function of $\phi_0$. Solid and dashed curves depict the positive and negative fermionic bound energies, respectively, in the dynamical model. Dotted lines depict the fermionic threshold energies in the background model. Right graph: The derivative of the fermionic negative bound energy with respect to $\phi_0$ as a function of $\phi_0$.}
    \label{fig:energy_spectrum}
\end{center}
\end{figure}

The left graph in figure \ref{fig:energy_spectrum} shows the fermionic bound state energies as a function of the asymptotic value of the bosonic field, $\phi_0$. As can be seen, the background result is retrieved as $\phi_0 \to \infty$, i.e.  $E=\pm\sqrt{3/2}\,(m)$. It is important to note that for $\phi_0\gtrsim2$ the dynamical graphs and the lines $E=\pm\sqrt{3/2}\,(m)$ are not easily distinguishable, although for smaller $\phi_0$ the negative and positive energies change drastically from the $\lambda\to0$ ($\phi_0 \to \infty$) limit result. 
In the numerical simulations the closer $\phi_0$ is to zero, the more difficult it is for the solutions to converge.
The smallest values of $\phi_0$ we are able to obtain are $\phi_0=0.501$ for the positive bound energy and $\phi_0=0.564$ for the negative one, though based on physical intuition it is possible to partially guess how the energy curves would behave below these values. 
The positive bound energy curve should not cross the zero energy line as it would configure level crossing \cite{farid}. 
Furthermore, as the negative energy curve becomes closer to the threshold line $E=-\sqrt{2}\,(m)$, its slope decreases considerably at $\phi_0\approx0.63$, as the right graph of figure \ref{fig:energy_spectrum} shows. 

\begin{figure}[h]
\begin{center}
	\fbox{\includegraphics[width=0.8\textwidth]{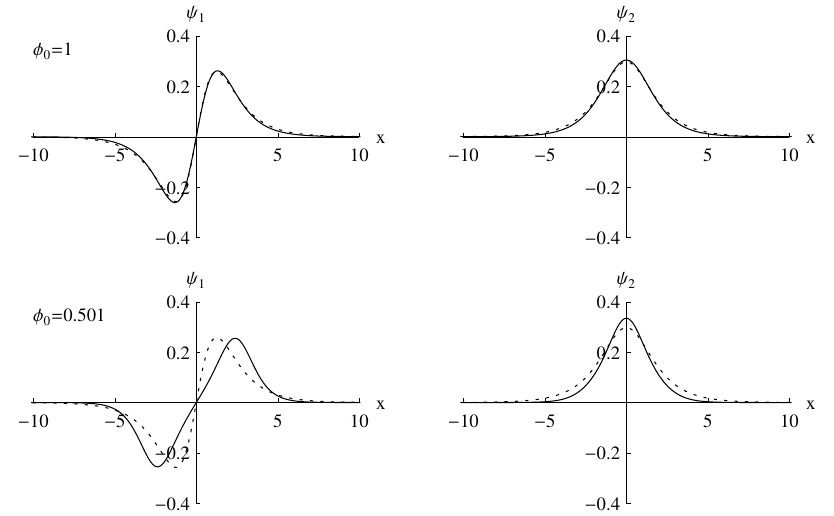}}
	\caption{The positive energy fermionic bound state as a function of $x$ for $\phi_0 = 1$ and $\phi_0 = 0.501$. Solid and dashed curves show the fermionic bound state in the dynamical and background models, respectively.}
    \label{fig:psi_wave_functions_positive}
\end{center}
\end{figure}

\begin{figure}[h]
\begin{center}
	\fbox{\includegraphics[width=0.8\textwidth]{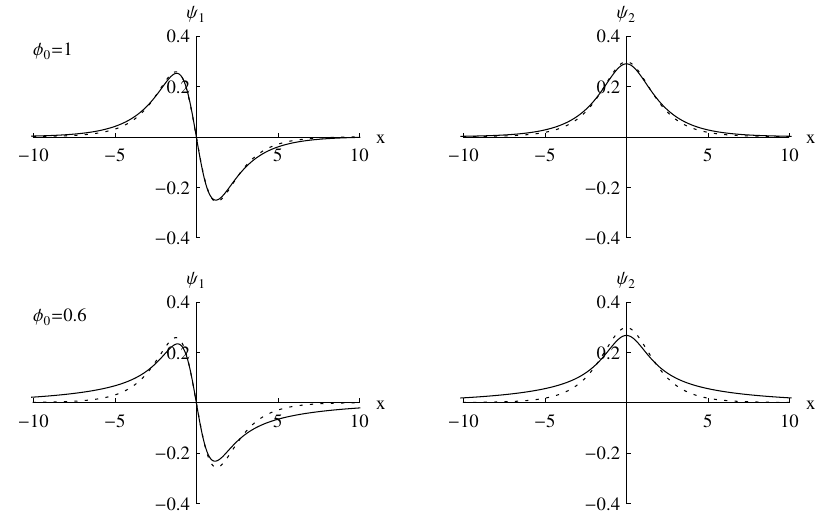}}
	\caption{The negative energy fermionic bound state as a function of $x$ for $\phi_0 = 1$ and $\phi_0 = 0.6$. Solid and dashed curves show the fermionic bound state in the dynamical and background models, respectively.}
    \label{fig:psi_wave_functions_negative}
\end{center}
\end{figure}

\begin{figure}[h]
\begin{center}
	\fbox{\includegraphics[width=0.8\textwidth]{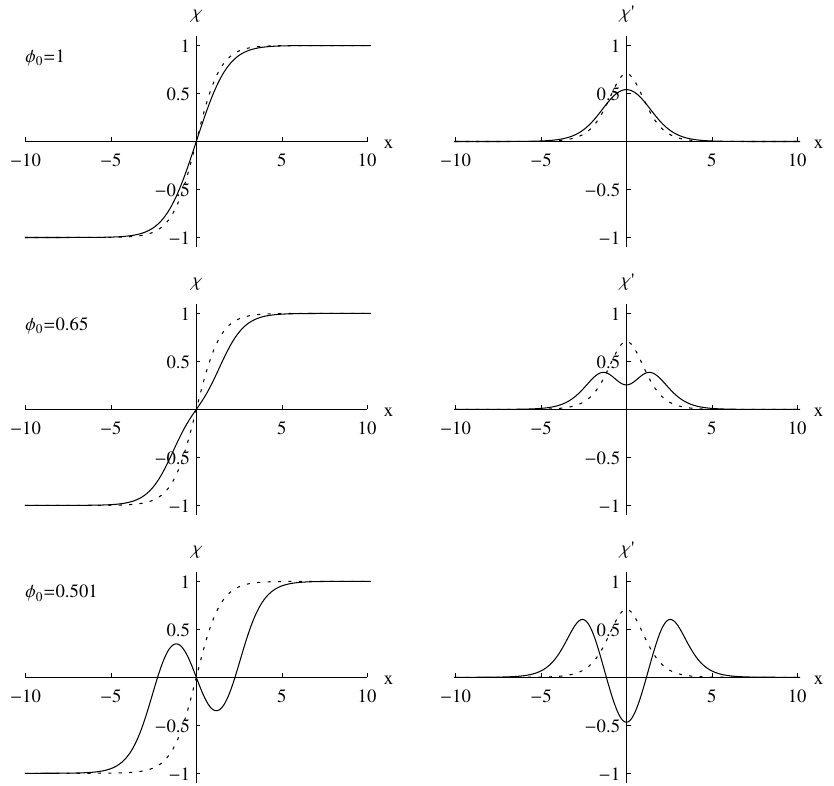}}
	\caption{The bosonic field and its derivative with respect to $x$ corresponding to positive bound energy for $\phi_0 = 1$, $\phi_0 = 0.65$ and $\phi_0 = 0.501$. Solid and dashed curves depict the bosonic field and its derivative in the dynamical and background models, respectively.}
    \label{fig:phi_wave_functions_positive}
\end{center}
\end{figure}

\begin{figure}[h]
\begin{center}
	\fbox{\includegraphics[width=0.8\textwidth]{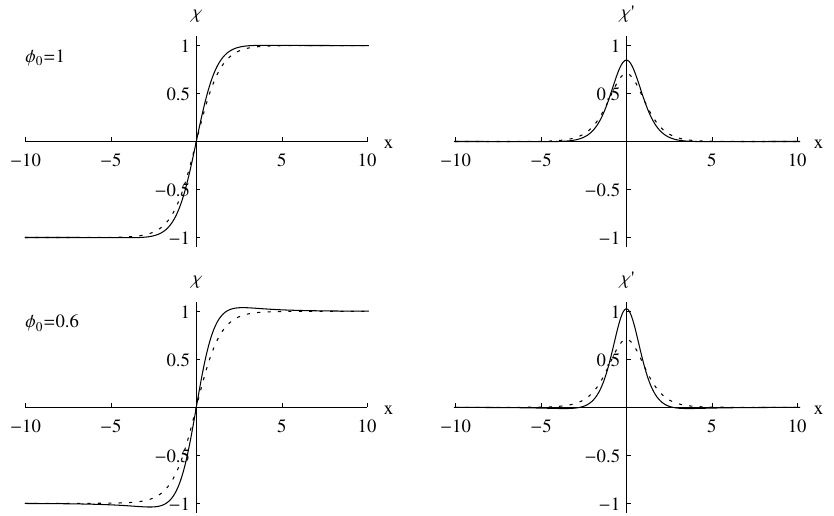}}
	\caption{The bosonic field and its derivative with respect to $x$ corresponding to negative energy for $\phi_0 = 1$ and $\phi_0 = 0.6$. Solid and dashed curves depict the bosonic field and its derivative in the dynamical and background models, respectively.}
    \label{fig:phi_wave_functions_negative}
\end{center}
\end{figure}

Figures \ref{fig:psi_wave_functions_positive} and \ref{fig:psi_wave_functions_negative} depict the fermionic bound states as a function of $x$ for positive and negative energies, respectively. The solid curves are the result of our dynamical model and the dotted ones are the result in $\lambda\to0$ limit. We show the results for two different low values of $\phi_0$ for positive and negative energy states in order to highlight the effects in the region far from $\lambda\to0$ limit. As can be seen in the graphs, in lower $\phi_0$ case the dynamical and background results become more distinct which confirms that in the low $\phi_0$ or large coupling $\lambda$ region the system cannot be described by the background approximation.
To investigate the effect of the fermion on the shape of the soliton, in Figures \ref{fig:phi_wave_functions_positive} and \ref{fig:phi_wave_functions_negative} we show the bosonic field as a function of $x$ for positive and negative energies, respectively.  
Since the results for positive energy change considerably in low $\phi_0$ region, we show the result for three distinct values of $\phi_0$ to make it possible to track the transition to the large coupling $\lambda$ limit. For each of the graphs we show $\chi$ and its spatial derivative, $\chi'$, to illustrate the way the soliton changes from the background one.
Interestingly enough, although the slope of the kink of $\lambda \phi^4$ theory is always positive, the interaction with the fermion can be strong enough to invert the sign of the slope at the origin.
It is important to notice that although the soliton can change drastically in the large $\lambda$ region, the changes are limited to a small region around the origin. This result shows that the back-reaction of the fermion on the soliton and thus the disturbance region are finite, except eventually for the limit $\phi_0\rightarrow 0$.

\begin{figure}[h]
\begin{center}
	\fbox{\includegraphics[width=0.5\textwidth]{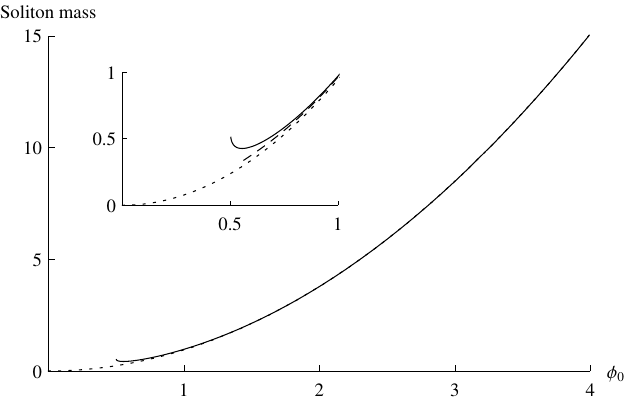}}
	\caption{The classical soliton mass as a function of $\phi_0$. Solid and dashed curves show the classical soliton mass in the dynamical model corresponding to positive and negative bound energies, respectively. Dotted line is the classical soliton mass for the background model.}
    \label{fig:soliton_mass}
\end{center}
\end{figure}

Using the expression (\ref{eq:soliton_mass}), the classical mass of the soliton  is shown in Fig.\,\ref{fig:soliton_mass} for both positive and negative energy bound states in the dynamical model as well as the background one. 
As can be seen for low $\phi_0$ the mass of the soliton diverges significantly from the one in $\lambda\to0$ limit. However, for $\phi_0$ greater than $1$ the three curves coincide within the scale shown in the graph. 

\begin{figure}[h]
\begin{center}
	\fbox{\includegraphics[width=0.75\textwidth]{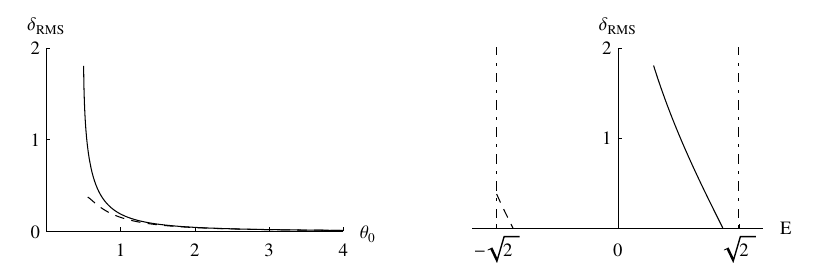}}
	\caption{The back-reaction of the fermion on the soliton, $\delta_{RMS}$, as a function of $\phi_0$ and the fermionic bound energy $E$. Solid and dashed curves show the back-reaction corresponding to the positive and negative bound energies, respectively, in the dynamical model. Dotdashed lines are the threshold energies in the background model.}
    \label{fig:backreaction}
\end{center}
\end{figure}

Again, as a measure of the effect of the fermion on the soliton we calculate the root mean squared deviation between the prescribed and dynamical soliton, $\delta_{RMS}$. In figure \ref{fig:backreaction}, we show this result as a function of $\phi_0$ and energy, for both positive and negative bound states. 
As expected, the back-reaction of the fermion on the soliton goes to zero when $E\to\pm\sqrt{3/2}\,(m)$, i.e. $\lambda\to0$ result. Interestingly, the right graph in this figure shows that the back-reaction decreases almost linearly with energy. Also, the left graph of this figure confirms that when $\phi_0$ goes to zero the back-reaction increases significantly and cannot be neglected in this region, i.e. the large coupling region.

\section{Conclusion}
In this paper we have investigated a fermion-soliton model in $(1+1)$ dimensions in which a static pseudoscalar field interacts nonlinearly with a Dirac particle. In this system the bosonic self-interaction part of the potential that is responsible for creating a soliton with proper topological characteristics has been considered to be the potential in $\lambda \phi^4$ theory.
First we have considered the general case where the Yukawa coupling, $g$, is independent from the bosonic self-coupling, $\lambda$. 
Then, we have solved a minimal supersymmetric kink model which is a particular example of the former with $g = \sqrt{\lambda/2}$.

We have found the zero mode fermionic state and threshold energies analytically, although in order to find other bound states and the corresponding shape of the soliton we needed a numerical method. 
We used a relaxation method to calculate the energy spectrum and the bound states as well as the shape of the soliton. In the general case, where the couplings $g$ and $\lambda$ are independent, we have solved the system for the soliton with winding number one, i.e. $\phi_0 = \pi$. As a consistency check, we have studied the limit where the soliton can be considered as background, $g/\phi_0 \to 0$ ($\lambda \to 0$ in supersymmetric case). 

Our calculations have shown that the back-reaction of the fermion on the soliton for the fermionic zero mode is zero. 
Therefore, the soliton corresponding to the fermionic zero mode is the kink of $\lambda \phi^4$ theory, even in the case where $g/\phi_0$ ($\lambda$ in supersymmetric case) is large.
However, except in $g/\phi_0 \to 0$ limit, the system does not have particle conjugation symmetry which would guarantee that the soliton receives no back-reaction from the fermionic zero mode. This happens because the term $g\bar \psi\psi$ is zero in the zero mode case and the system retrieves the particle conjugation symmetry. 
Besides that, since the particle conjugation symmetry is broken for finite $g/\phi_0$, the energy spectrum becomes progressively asymmetric around $E=0$ with increasing $g/\phi_0$.

Our numerical results have shown that the energy spectrum converges to the result of the background model as $g/\phi_0\to0$, unsurprisingly.
The same happens with the classical soliton mass. However, they are completely distinguishable when $g/\phi_0$ is large.
In the supersymmetric case by varying the value of $\phi_0$ from zero to infinity we could span the region between $g/\phi_0\to0$ and large $g/\phi_0$, within the numerical limitations. 

Furthermore, we have calculated the back-reaction of the fermion on the soliton for the positive and negative energy states as a function of $\phi_0$ and $E$ for both general Jackiw-Rebbi model and the supersymmetric case.
The results show that the back-reaction of the fermion on the soliton tends to zero as $g/\phi_0\to0$ for both positive and negative bound energy curves, as expected. In contrast, with large $g/\phi_0$ the back-reaction increases significantly. 
When the value of $g/\phi_0$ is high enough it can distort the shape of the soliton to the point that the slope of the soliton at the origin becomes negative, even though for the kink the slope is always positive.
Therefore, the background soliton approximation can fail drastically for large $g/\phi_0$.

\section*{Acknowledgments}
A.M. and A.A. thank E. R. Bezerra de Mello for the helpful discussions.
The authors thank Conselho Nacional de Desenvolvimento Cient\'{\i}fico e Tecnol\'{o}gico (CNPq) for the financial support. Also, A.M. thanks PNPD/CAPES for the partial support.

\end{document}